\documentclass{PoS}

\title{{\vspace{-15mm} \normalsize\hfill{\small DESY 10-187}}\\[15mm]{\vspace{-18mm} \normalsize\hfill{\small SFB/CPP-10-98}}\\[10mm]Overlap Valence Quarks on a Twisted Mass Sea}

\ShortTitle{Overlap Valence Quarks on a Twisted Mass Sea}

\author{\speaker{K. Cichy}\thanks{Supported by Ministry of Science and Higher Education grant nr. N N202 237437. The computer time for this project was made available to us by the Leibniz Rechenzentrum in Munich, Pozna\'n Supercomputing and Networking Center in Pozna\'n, GENCI-CINES grant 2010-052271 and CCIN2P3 in Lyon. We thank these 
computer centers and their staff for all technical advice and help.}\\
        Adam Mickiewicz University, Faculty of Physics, Umultowska 85, 61-614 Pozna\'n, Poland\\
        E-mail: \email{kcichy@amu.edu.pl}}

\author{V. Drach, E. Garcia-Ramos, G. Herdoiza, K. Jansen\\
        NIC, DESY, Platanenallee 6, D-15738 Zeuthen, Germany\\
}

\abstract{We present the results of an investigation of a mixed action approach of 
overlap valence and maximally twisted mass sea quarks. Employing a particular matching condition
on the pion mass, we analyze the continuum limit scaling of the pion decay 
constant and the role of chiral zero modes of the overlap operator in this process. 
We employ gauge field configurations generated by the European Twisted Mass 
Collaboration with linear lattice size $L$ ranging from 1.3 to 1.9 fm. The 
continuum limit is taken at a fixed value of $L=1.3$ fm, employing three 
values of the lattice spacing and two values of the pion mass constructed 
from sea quarks only.
}

\FullConference{The XXVIII International Symposium on Lattice Field Theory, Lattice2010\\
		June 14-19, 2010\\
		Villasimius, Italy}

\begin{document}

\section{Introduction}
Mixed action simulations with overlap fermions are considered to be a 
cost-effective alternative to fully dynamical overlap simulations. They allow to 
profit from the good chiral properties of overlap fermions in the valence sector, 
while in the sea sector a cheaper fermion discretization is used to keep the 
simulation time to a tolerable level.

We have used several ensembles of gauge field configurations generated by the 
European Twisted Mass Collaboration (ETMC) to investigate the particular 
setup of overlap valence quarks on a maximally twisted mass sea. Our aim was to 
explore the potential of this approach and study the continuum limit behaviour
of this mixed action setup. 
For earlier results 
regarding this setup, we refer to \cite{OVonTM1,OVonTM2,zakopane09}.


\section{Simulation setup}
We use ETMC $N_f=2$ dynamical maximally twisted mass (MTM) gauge field 
configurations and we refer to
refs.~\cite{boucaud,boucaud2}  
for the details concerning their generation.             
The parameter values are gathered in Tab.~\ref{tab-param}. 
We focus mainly on small-volume ensembles with $L\approx1.3$ fm, with pion 
masses corresponding to about 300 and 450 MeV, but we also consider larger 
volumes to investigate the size of finite volume effects in the case of the coarsest lattice spacing.

\begin{table}
\begin{center}
\vspace{0.5cm}
\begin{tabular}[c]{|c|c|c|c|c|c|c|}
\hline
Lattice size & $L$ [fm] & $\beta$ & $a$ [fm] & $a\mu$ & $m_\pi$ [MeV] & \#conf\\
\hline\hline
$16^3\times32$ & 1.3 & 3.9 & 0.079 & 0.004 & 300 & 544\\
$20^3\times40$ & 1.3 & 4.05 & 0.063 & 0.003 & 300 & 300\\
$24^3\times48$ & 1.3 & 4.2 & 0.051 & 0.002 & 300 & 401 \\
$16^3\times32$ & 1.3 & 3.9 & 0.079 & 0.0074 & 450 & 260 \\
$20^3\times40$ & 1.3 & 4.05 & 0.063 & 0.006 & 450 & 299 \\
$24^3\times48$ & 1.3 & 4.2 & 0.051 & 0.005 & 450 & 137 \\
$20^3\times40$ & 1.6 & 3.9 & 0.079 & 0.004 & 300 & 239\\
$24^3\times48$ & 1.9 & 3.9 & 0.079 & 0.004 & 300 & 435\\
\hline
\end{tabular}
\caption{We give the simulation parameters and indicate the values of $L$, $a$ and 
$m_\pi$ to provide estimates for the physical situation.}
\label{tab-param}
\end{center}
\end{table}

In the valence sector, we use overlap fermions \cite{neuberger97}, defined by:
\begin{equation}
\label{massless_overlap}
\hat D_{\rm overlap}(0)=\frac{1}{a}\Big(1-A(A^\dagger A)^{-1/2}\Big).
\end{equation}
with the kernel operator $A=1+s-a\hat D_{\rm Wilson}(0)$,
where $s$ is a parameter which satisfies $|s|<1$ and can be used to optimize 
locality properties and the Wilson-Dirac operator is defined by:
\begin{equation}
\hat D_{\rm Wilson}(m)=\frac{1}{2}\left(\gamma_\mu(\nabla_\mu^*+\nabla_\mu)-ar\nabla_\mu^*\nabla_\mu\right)+m,
\end{equation}
where $m$ is the bare Wilson quark mass and $\nabla_\mu$, $\nabla^*_\mu$ are the 
forward and backward covariant derivatives, respectively.
The massive overlap Dirac operator is given by:
\begin{equation}
\label{overlap-massive}
\hat D_{\rm overlap}(m_q) = \left(1+s-\frac{am_q}{2}\right)\hat D_{\rm overlap}(0)+m_q,
\end{equation} 
where $m_q$ is the bare overlap quark mass.

For each ensemble, following ref.~\cite{sumr} 
we use the SUMR solver with adaptive precision and multi mass 
capability to compute the propagators for a wide range of valence quark masses 
$m_q$, from quark mass matched to the unitary one to around the mass of the 
physical strange quark. Before applying the overlap operator, we perform one 
iteration of HYP smearing \cite{hyp} on the gauge field configurations.

\section{Locality of the overlap operator}
In order to achieve the best locality properties of the overlap operator, 
the parameter $s$ can be tuned. It has been shown \cite{hernandez} that 
the overlap operator is local for a wide range of simulation parameter values, i.e. 
the norm of the overlap operator falls off exponentially 
$||D_{overlap}||_{max}(d)\propto e^{-\rho d}$, where $d$ is the 
taxi-driver distance between lattice points and $\|.\|_\mathrm{max}$ 
is the same norm as used in ref.~\cite{hernandez}. We have computed 
the decay rate $\rho$ for several values of the parameter $s$, both for 
the overlap operator constructed on HYP smeared and on original,
unsmeared configurations.

\begin{figure}[t!]
\begin{center}
\includegraphics[width=0.4\textwidth,angle=0]{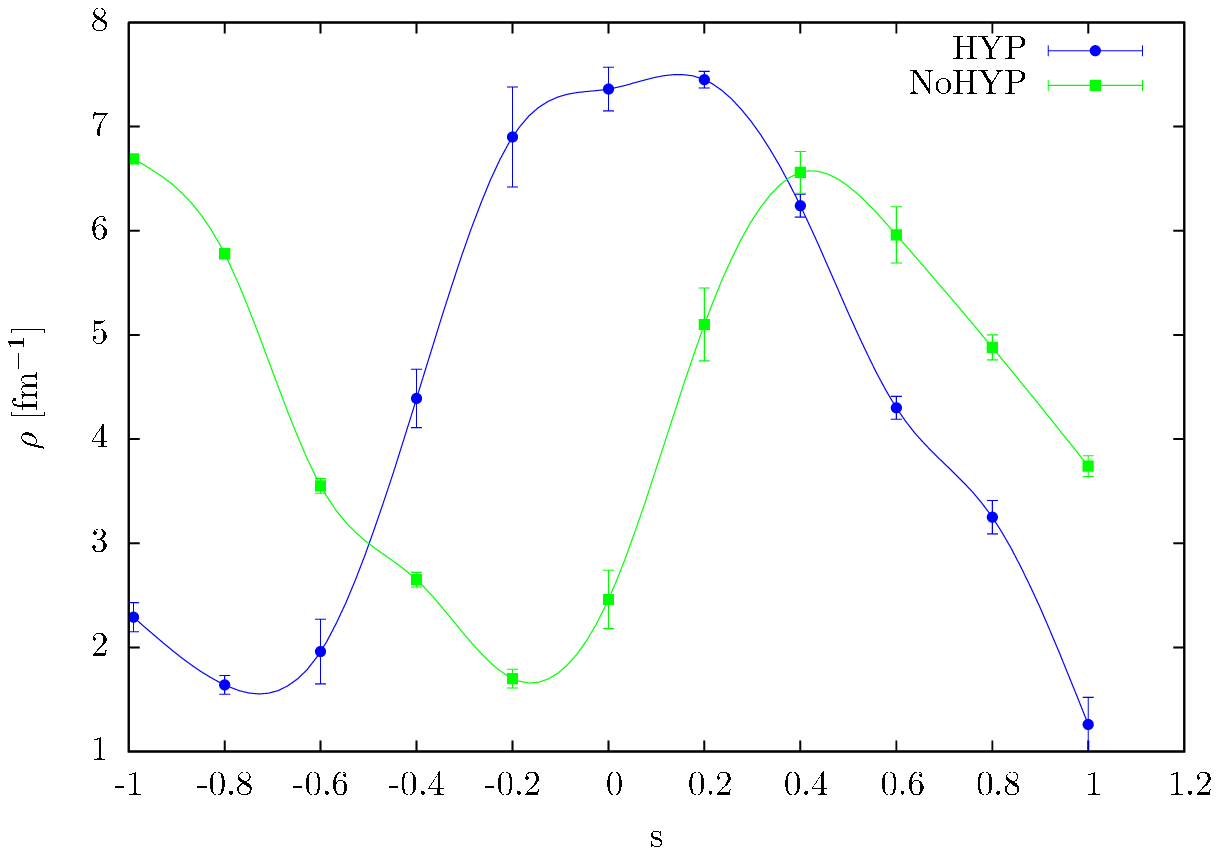}
\includegraphics[width=0.4\textwidth,angle=0]{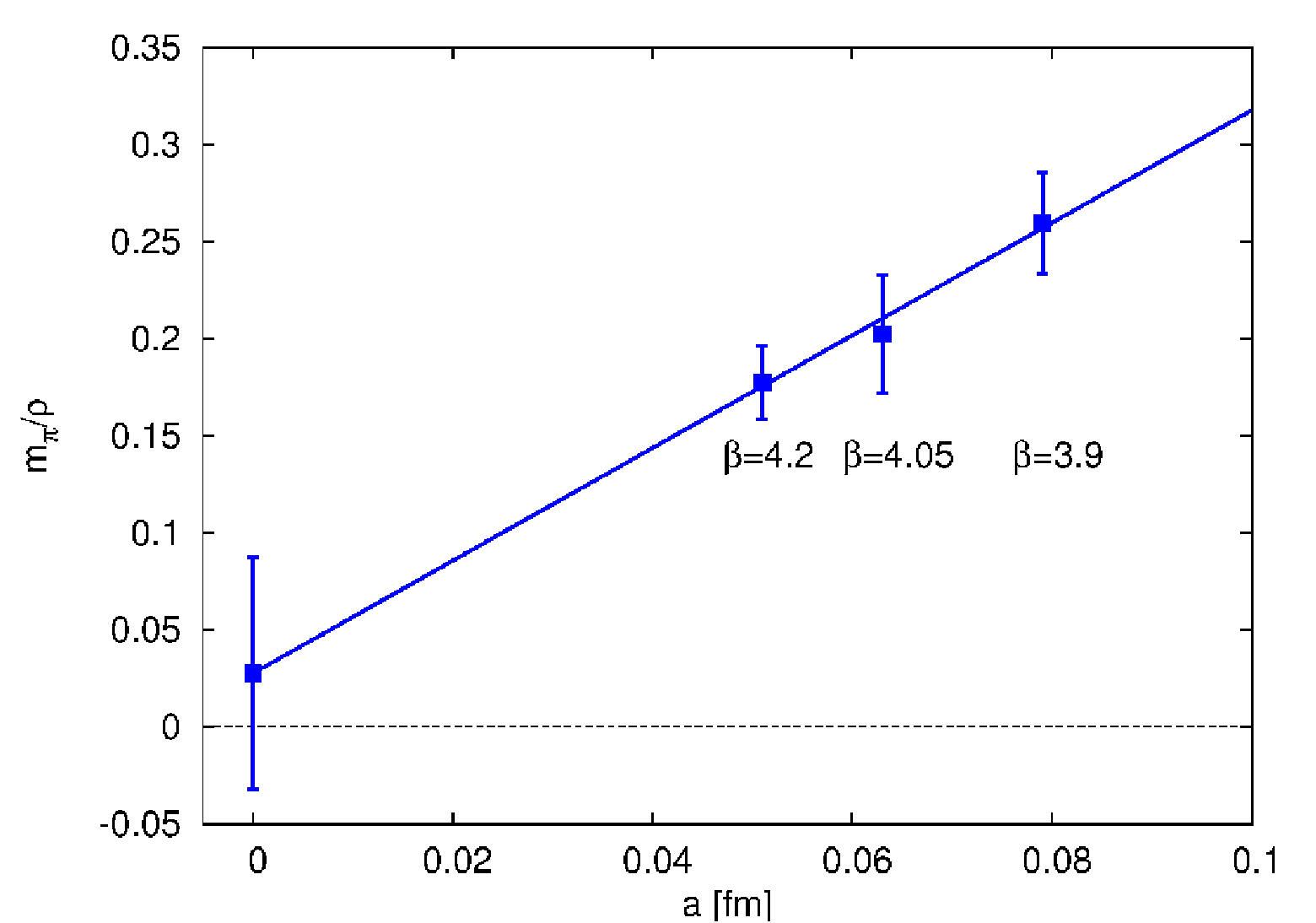}
\caption{(left) Decay rate of the overlap operator. (right) Continuum limit scaling of the ratio $m_\pi/\rho$.}
\label{fig-locality}
\end{center}
\end{figure}

We find (the left panel of Fig. \ref{fig-locality}) that in the HYP-smeared case the 
best locality is observed in the neighbourhood of the free-field optimal 
value $s=0$. Following this observation, we set $s=0$ in our computations. 
We also investigate the continuum limit of the ratio $m_\pi/\rho$. At finite lattice 
spacing the condition $m_\pi<\rho$ must hold \cite{niedermayer98} in order that 
the interaction can be considered local from the point of view of the 
considered particle. For the pion, the ratio $m_\pi/\rho$ 
(right panel of Fig. \ref{fig-locality}) is well below 1 and thus locality is 
guaranteed. Moreover, the continuum limit value of $m_\pi/\rho$ is consistent 
with zero, which is related to the fact that $m_\pi$ had been fixed to
$m_\pi\approx 300$MeV and has a non-zero 
value in the continuum, while $1/\rho$ vanishes for $a=0$.

\section{Continuum limit scaling test of the pion decay constant}
\begin{figure}[t!]
\begin{center}
\includegraphics[width=0.28\textwidth,angle=270]{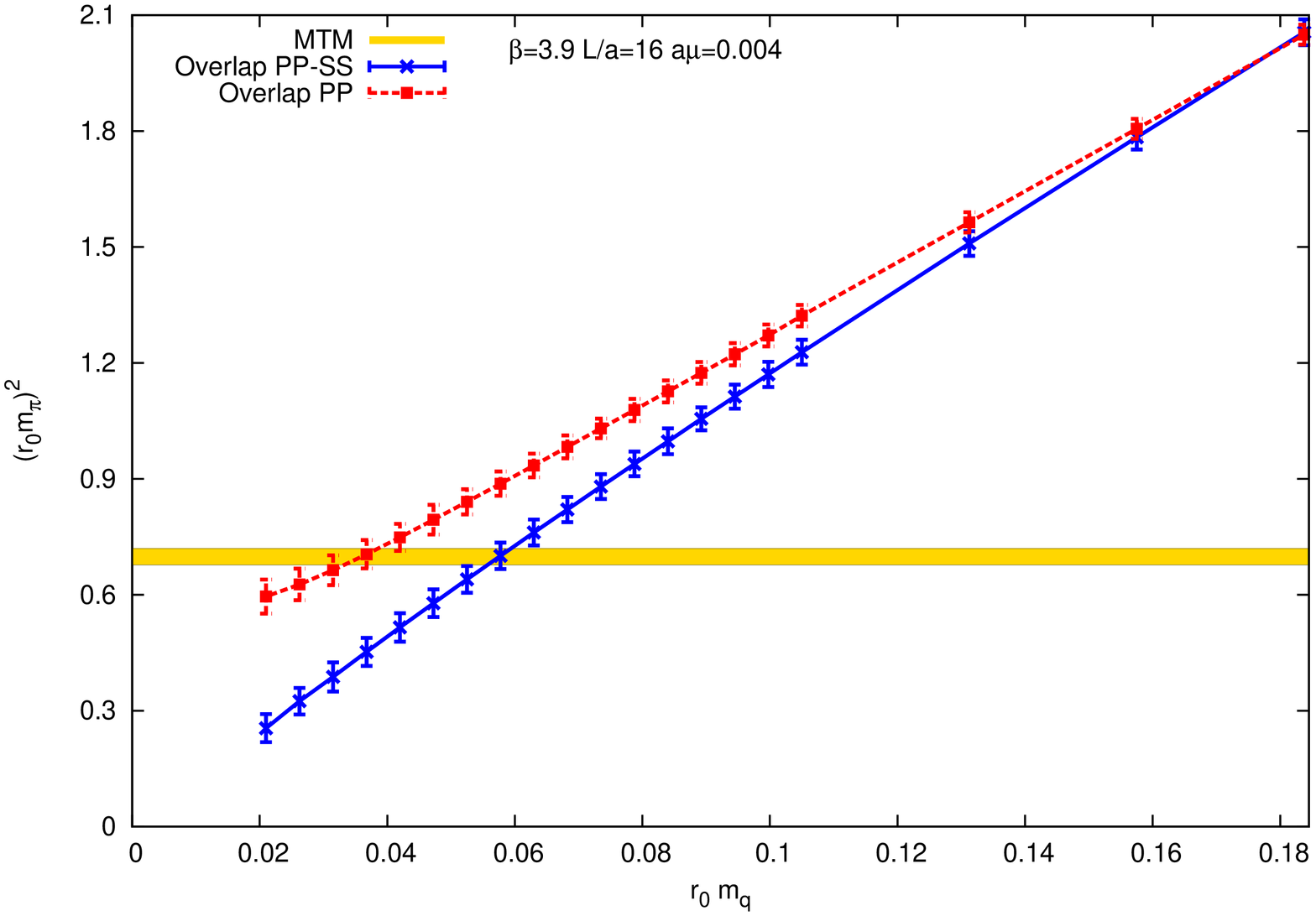}
\includegraphics[width=0.28\textwidth,angle=270]{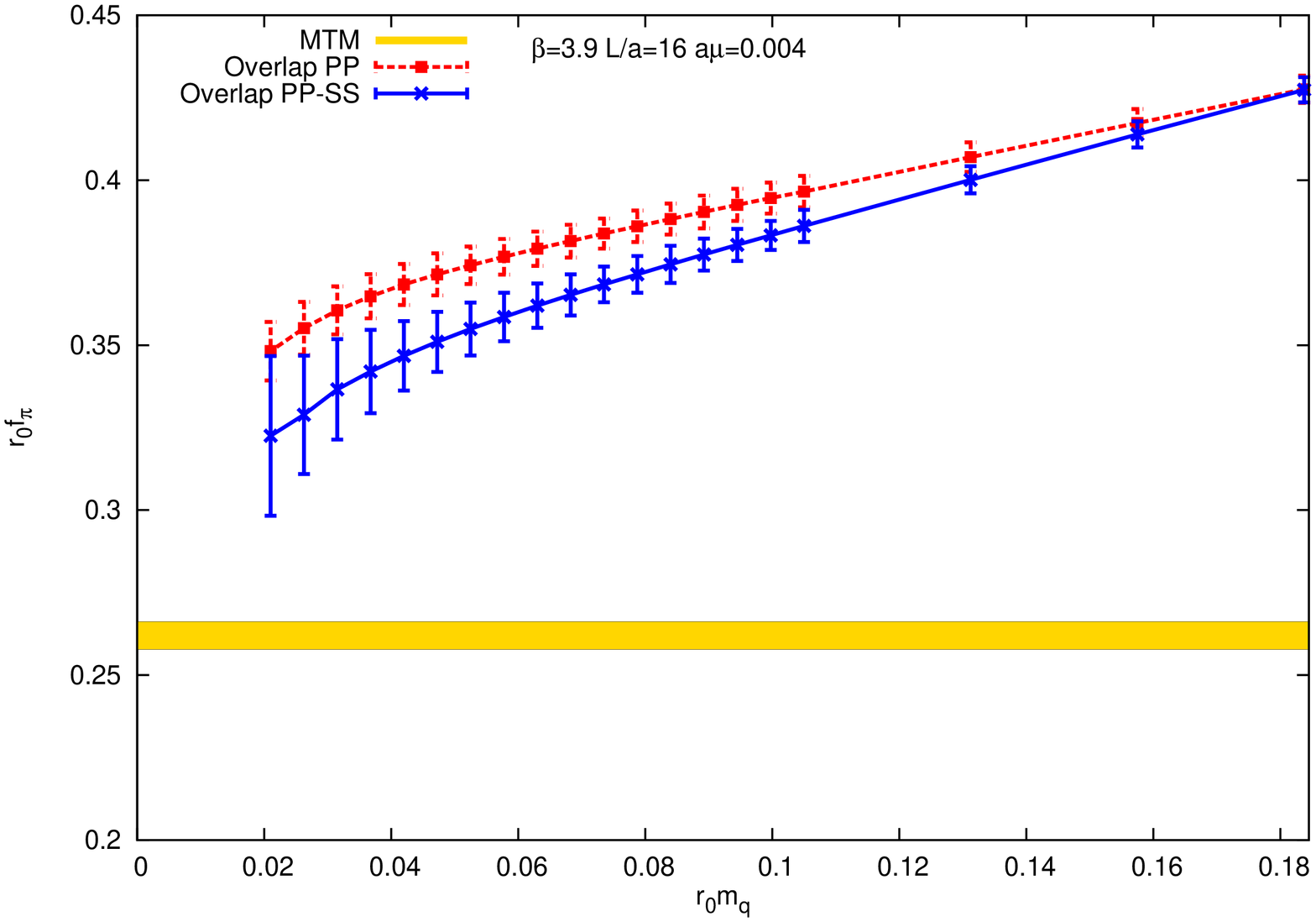}
\caption{The quark mass dependence of the pion mass (left) and the pion decay constant (right). Ensemble: $\beta=3.9$, $L/a=16$, $m_\pi\approx300$ MeV.}
\label{fig-massdep}
\end{center}
\end{figure}

To minimize the unitarity violations present in the mixed action setup, a procedure of matching of the quark masses has to be performed. 
In general, there is a large freedom to choose a matching condition. In 
this proceeding, we will discuss a particular matching condition which 
will illustrate the role of chiral zero modes of the overlap operator 
especially. Our condition consists of matching the value of the charged 
pion mass\footnote{In the twisted mass formalism, the neutral and charged pions have unequal masses, which is due to $\mathcal{O}(a^2)$ isospin symmetry breaking. We choose to match the charged pion mass, because of its smaller statistical uncertainty and smaller discretization errors.}, i.e. to enforce that the pion built from two valence quarks has 
a mass closest to the one built from two sea quarks.
The procedure of matching is illustrated in the left panel of Fig. \ref{fig-massdep}, which depicts the quark mass dependence of the pion mass. We extract the pion mass from two correlators -- the pseudoscalar one (PP) and the difference of the pseudoscalar and the scalar ones (PP-SS). Taking the latter, one can avoid any effects from the chiral zero modes of the overlap operator, since the zero modes couple equally to the PP and SS correlators. Thus, we calculate two matching quark masses for each ensemble.

It is clear that the effects of zero modes are non-negligible for this 
ensemble and the two definitions of the matching mass lead to different values of the
matching quark masses. Hence, we will use both definitions and analyze 
the effects at both matching quark masses.

The right panel of Fig. \ref{fig-massdep} shows the quark mass dependence of the pion decay constant $f_\pi$, where this observable has been extracted again from both the PP and the PP-SS correlator. The two curves tend to each other for large values of the quark mass, where the effects of the zero modes are small, but for small quark masses (including the matching mass) these effects are rather large.

\begin{figure}[]
\begin{center}
\includegraphics[width=0.28\textwidth,angle=270]{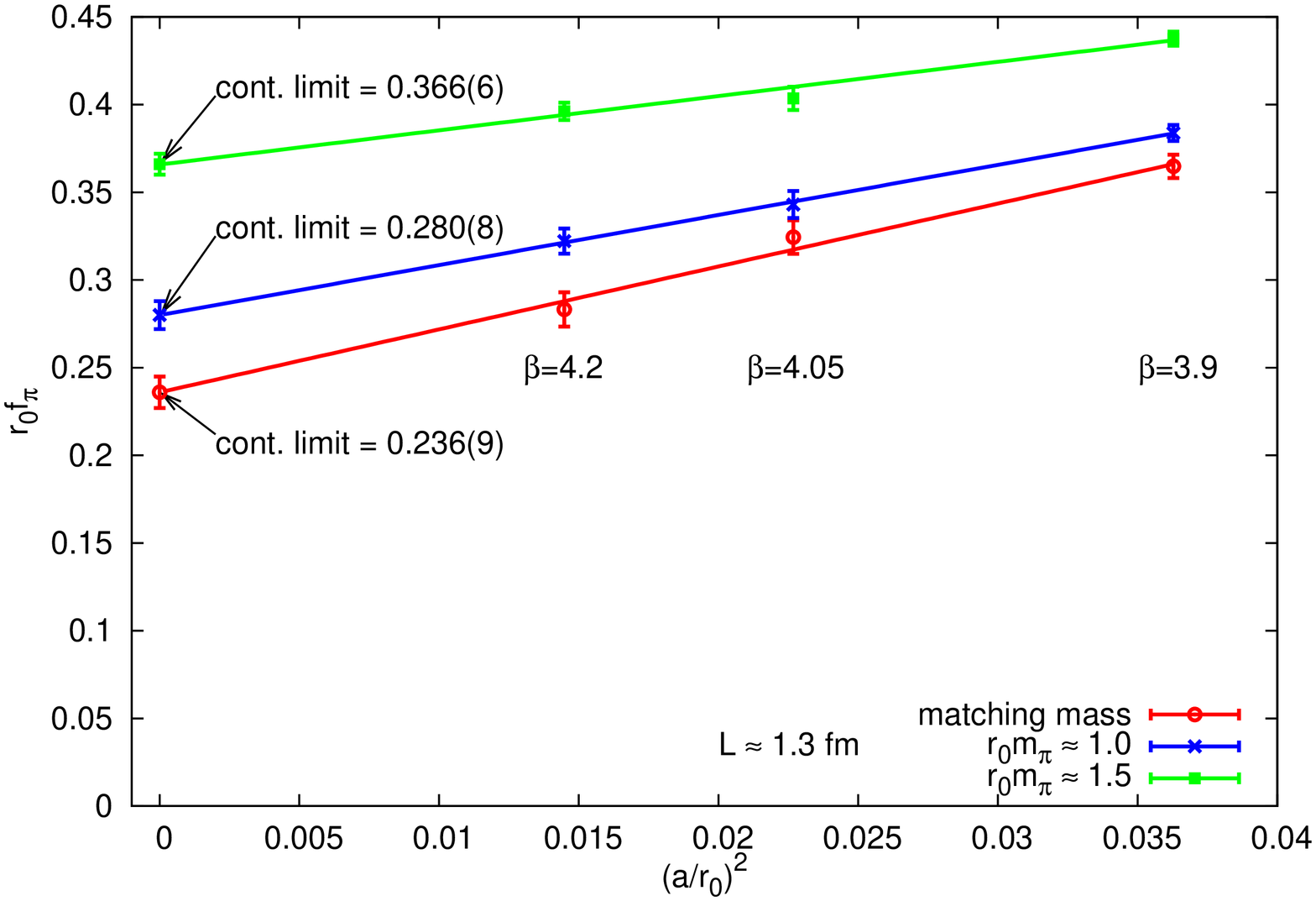}
\includegraphics[width=0.28\textwidth,angle=270]{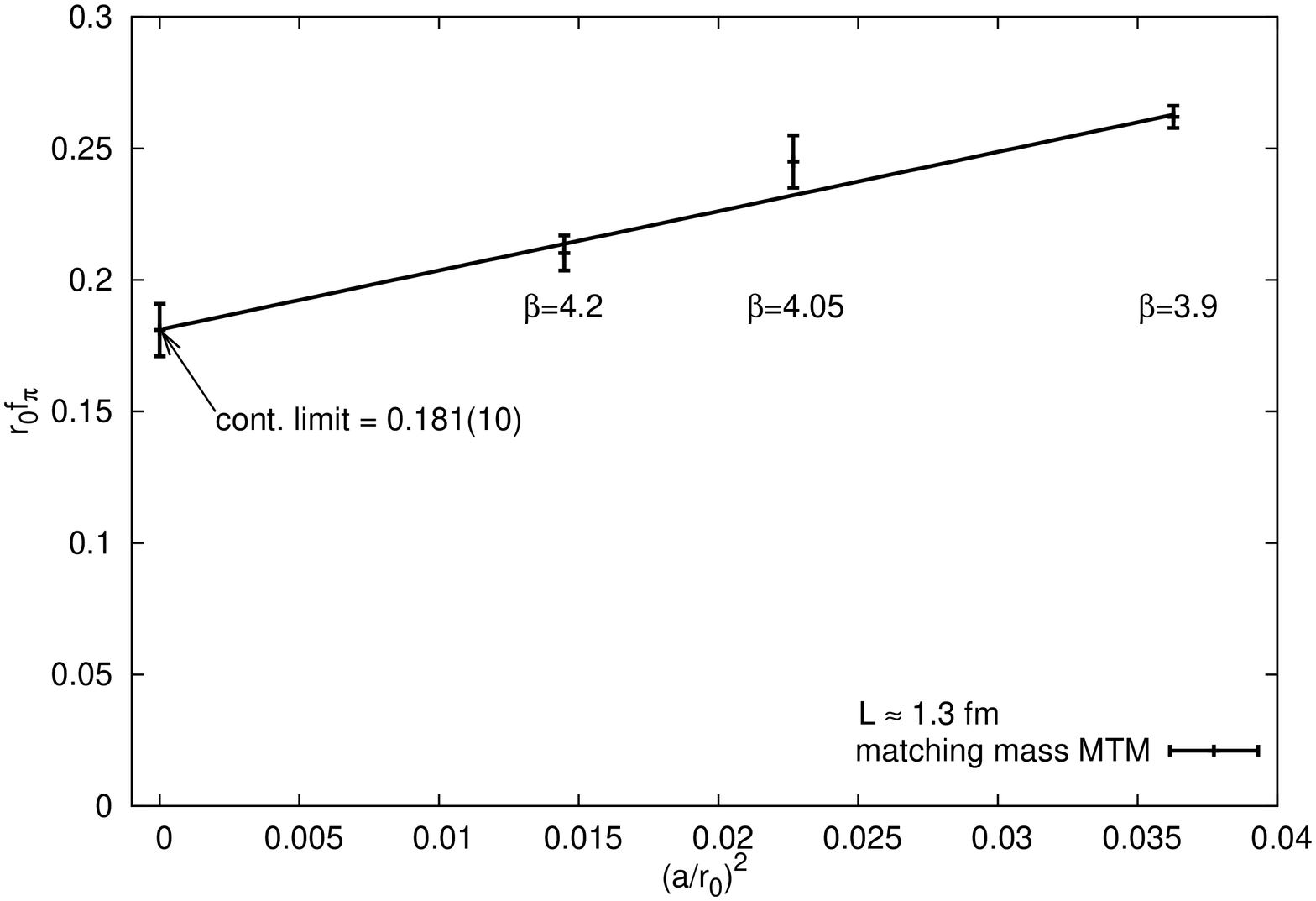}
\caption{The continuum limit scaling of the pion decay constant (extracted from the PP correlator) -- overlap pion at the matching mass and two other reference values of $r_0 m_\pi$ (left) and MTM pion (right).}
\label{fig-scaling}
\end{center}
\end{figure}

In the following, we concentrate on the continuum limit scaling test of the pion decay constant. We will perform it at three values of the pion mass, corresponding to $r_0m_\pi\approx0.85,1.0,1.5$, where the smallest value corresponds to the matching mass. Fig. \ref{fig-scaling} shows the scaling of $f_\pi$ (extracted from the PP correlator) for overlap valence quarks (left) and MTM valence quarks (unitary setup; right). Clearly, both discretizations show $\mathcal{O}(a^2)$ leading discretization effects.

\begin{figure}[]
\begin{center}
\includegraphics[width=0.28\textwidth,angle=270]{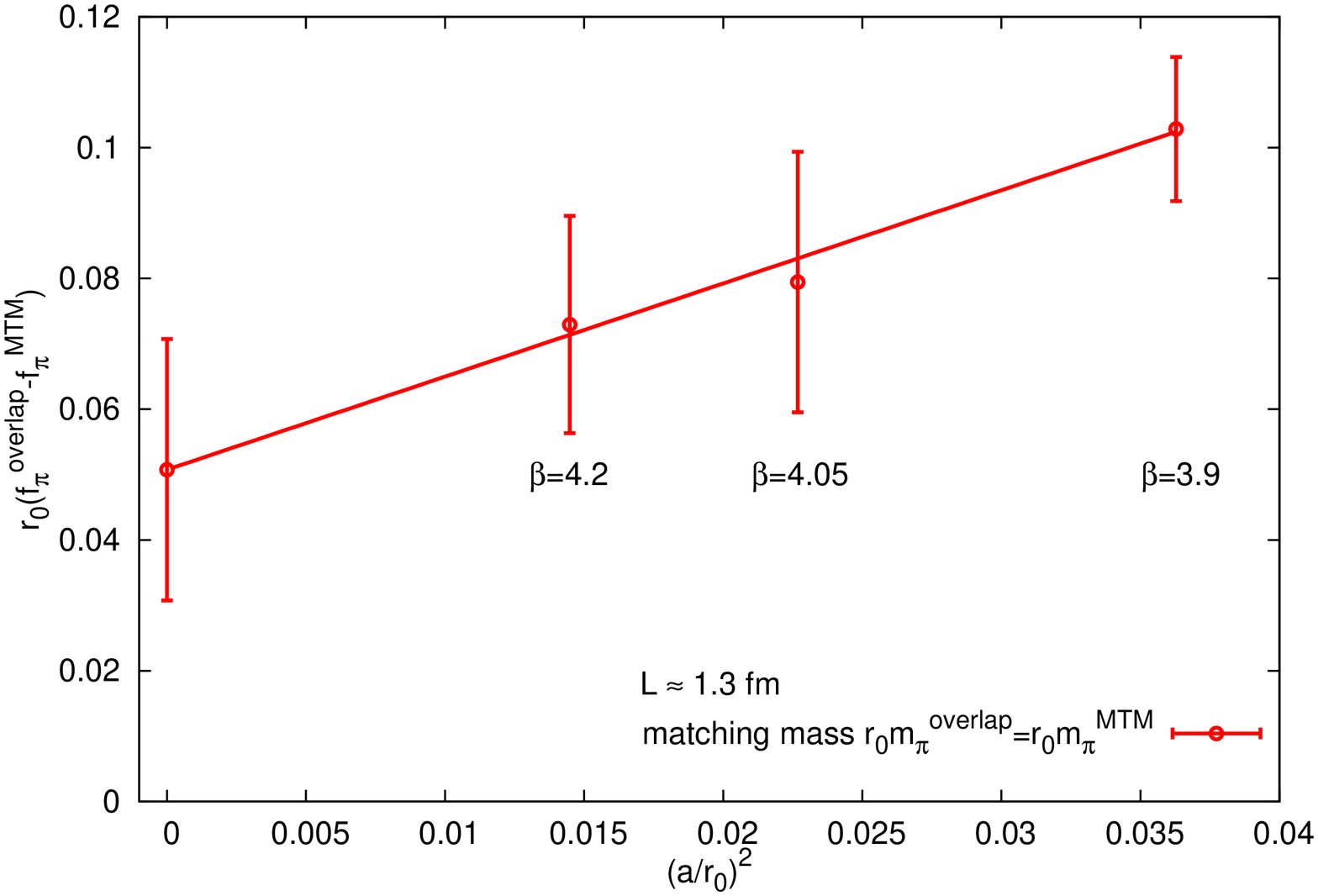}
\includegraphics[width=0.28\textwidth,angle=270]{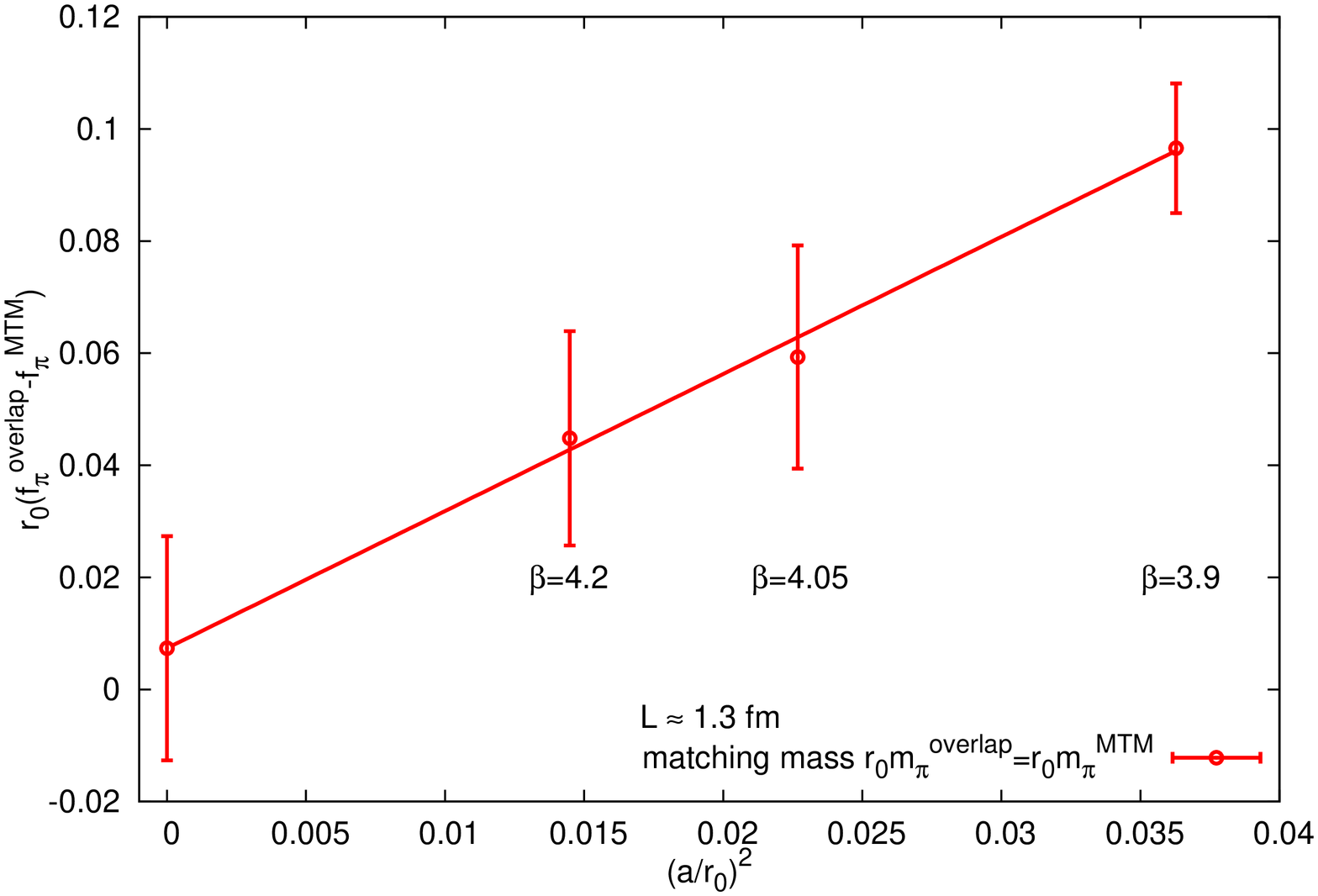}
\caption{The continuum limit scaling of the difference $r_0(f_\pi^{overlap}-f_\pi^{MTM})$, where $f_\pi^{overlap}$ has been extracted from the PP (left) and the PP-SS correlator (right).}
\label{fig-diff}
\end{center}
\end{figure}

However, when employing the matching condition discussed above, the 
continuum limit of $f_\pi$ in the mixed action setup is different 
from the unitary one, if $f_\pi$ is extracted from the PP correlator 
(left panel of Fig. \ref{fig-diff}) at the values of the lattice spacing
used here. This can be attributed to the effects of the zero modes, 
since the continuum limit value of $f_\pi$ extracted from the PP-SS correlator 
agrees nicely with the unitary value (Fig. \ref{fig-diff}, right).
We note that the seemingly inconsistent continuum limits originate from 
the particular matching condition used here. Employing alternative 
matching conditions can substantially alter this effect \cite{publ,publ2}.

\begin{figure}[]
\begin{center}
\includegraphics[width=0.28\textwidth,angle=270]{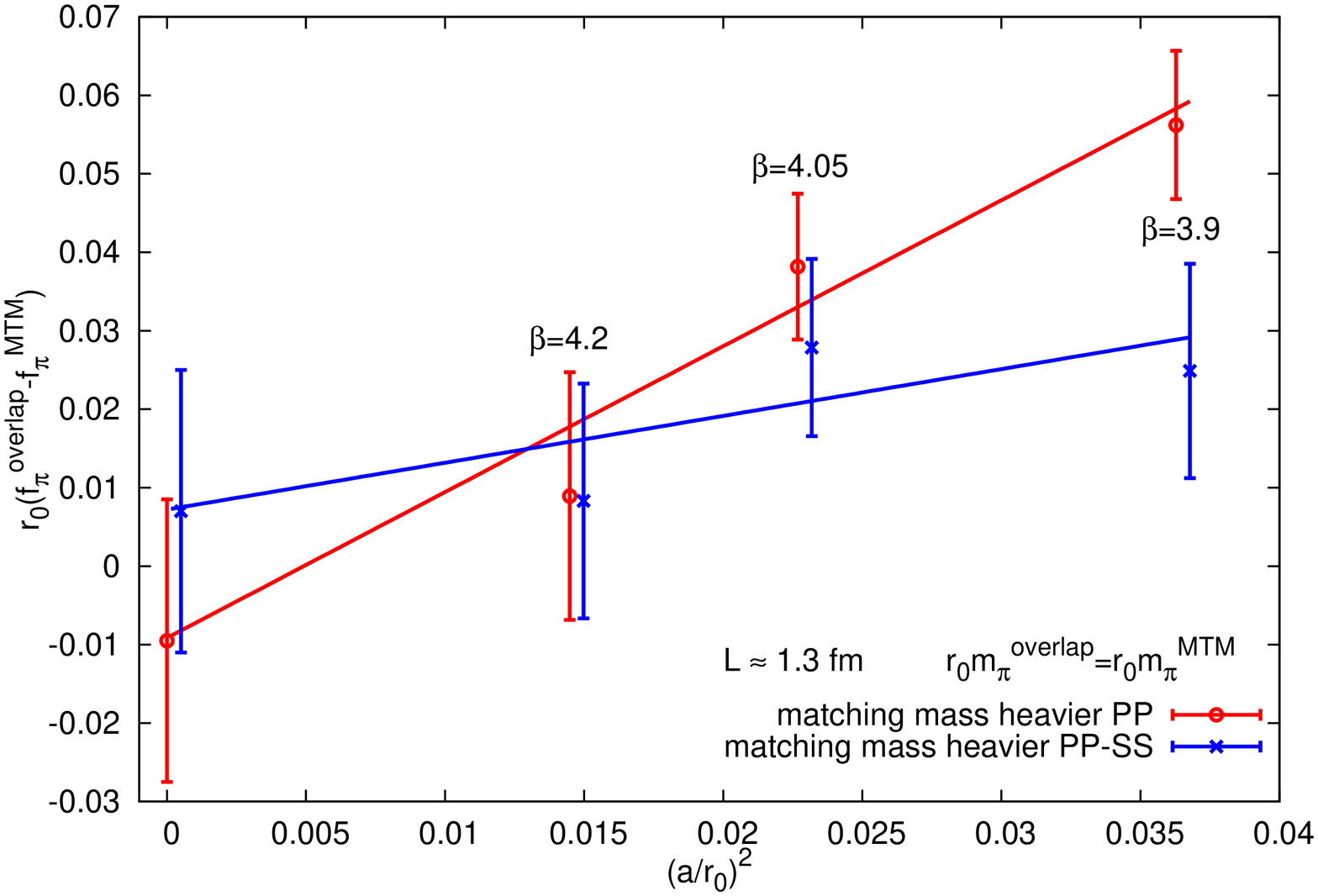}
\includegraphics[width=0.28\textwidth,angle=270]{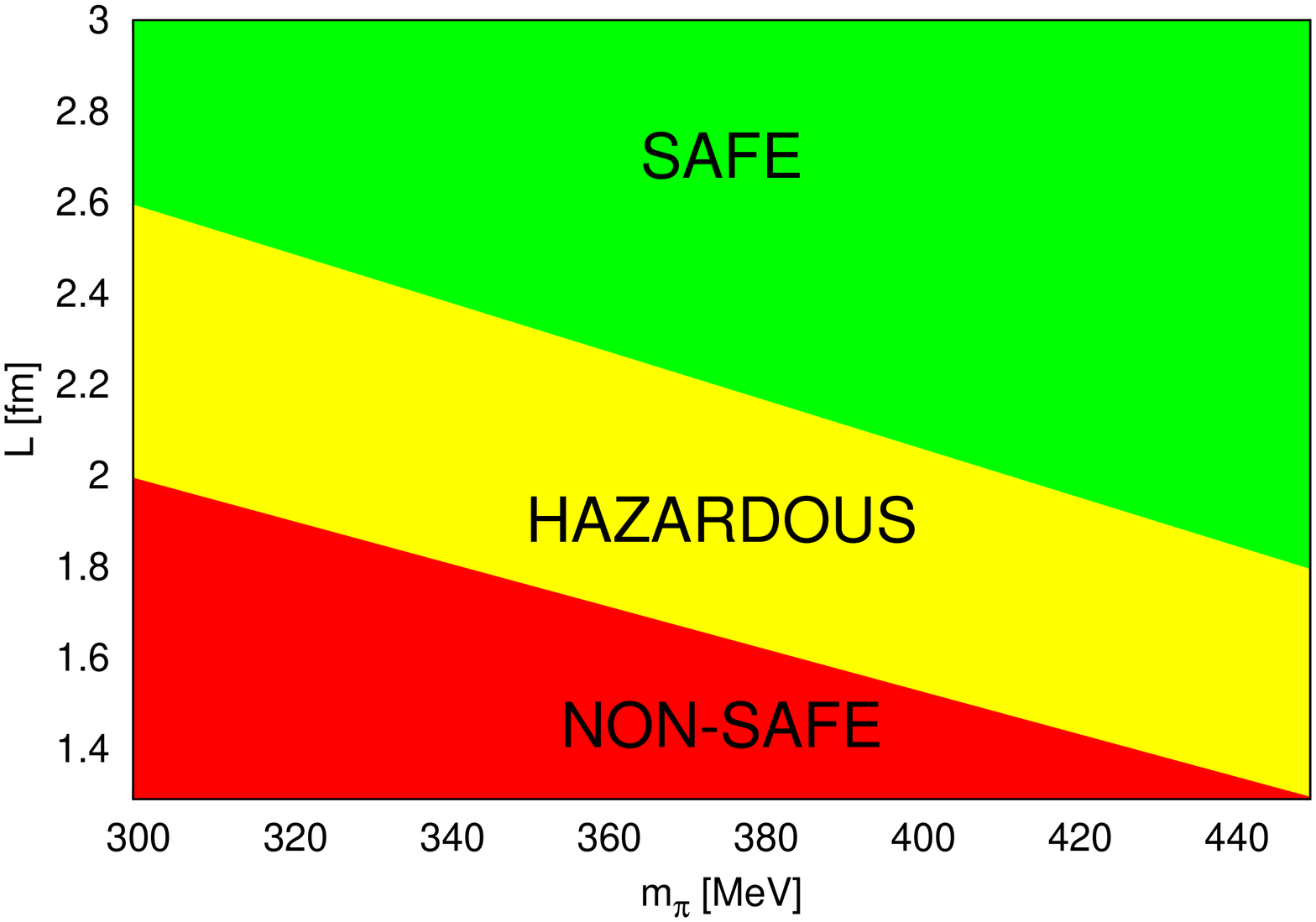}
\caption{(left) The continuum limit scaling of the difference $r_0(f_\pi^{overlap}-f_\pi^{MTM})$ in the case of a heavier sea quark mass ($m_\pi\approx450$ MeV). (right) The safe, hazardous and non-safe regions of parameter space with respect to the zero modes effects.}
\label{fig-heavier_safe}
\end{center}
\end{figure}

The hypothesis about the role of the zero modes can be confirmed by performing 
an analogous scaling test for a heavier value of the sea quark mass 
(corresponding to a pion mass of about 450 MeV). In this case, the zero modes 
effects (which enter the PP correlator as $1/m_q^2$ and $1/m_q$ effects) 
should be reduced. Indeed, the left panel of Fig. \ref{fig-heavier_safe} shows 
that the mixed action and unitary values of $f_\pi$ agree in the continuum limit, 
both in the PP and the PP-SS case. The effects of the zero modes are still visible, 
since $f_\pi$ extracted from PP and from PP-SS slightly differ, but these 
effects are strongly suppressed with respect to the $m_\pi\approx300$ MeV case.
Hence, when using a mixed action setup with overlap valence quarks, 
special attention has to be paid to the effects of the zero modes. 

The investigation of finite volume effects at the coarsest lattice spacing, together with the analysis of the sea quark mass dependence at small volume, allowed us to find the range of parameter values that ensure that one is safe against the effects of zero modes. This is summarized in the right panel of Fig. \ref{fig-heavier_safe}. In terms of $m_\pi L$ the safe region corresponds to $m_\pi L>4$ and the non-safe one to $m_\pi L<3$. For the details of this analysis, we refer to an upcoming publication \cite{publ}.

\section{Light baryon masses}
\begin{figure}[]
\begin{center}
\includegraphics[width=0.28\textwidth,angle=270]{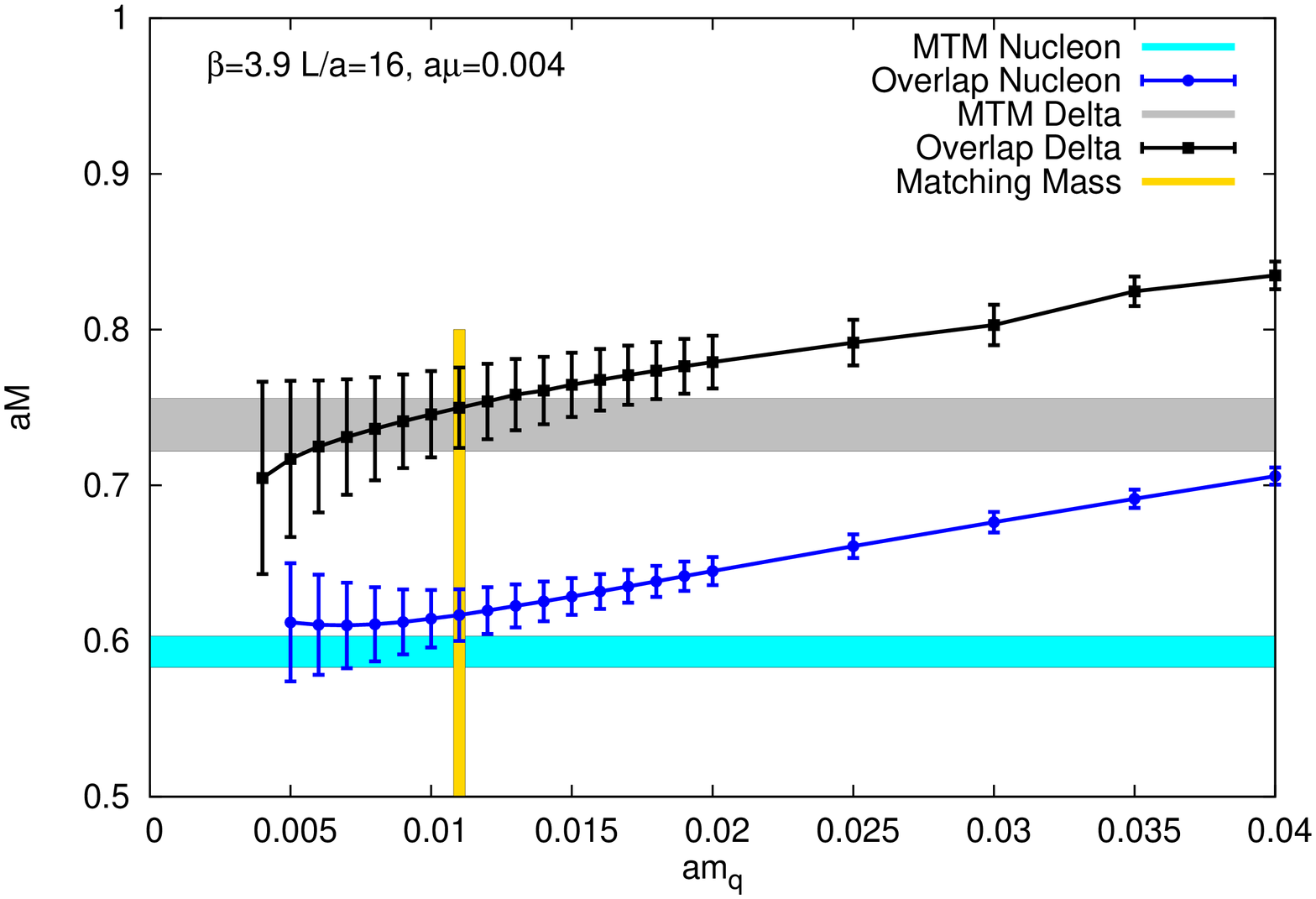}
\includegraphics[width=0.28\textwidth,angle=270]{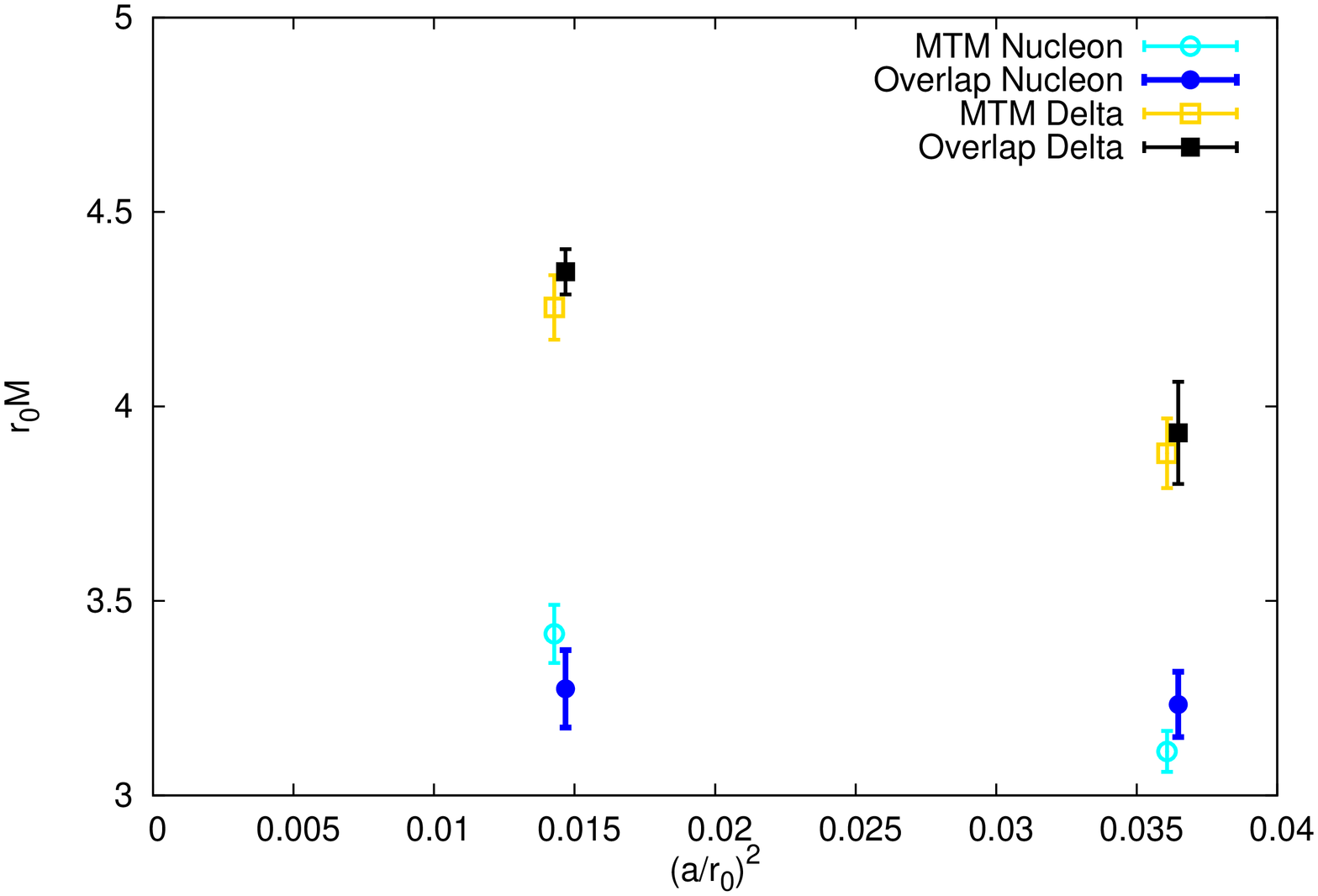}
\caption{(left) The quark mass dependence of the nucleon and delta mass. Ensemble: $\beta=3.9$, $L/a=16$, $a\mu=0.004$. (right) The unitary (MTM) and mixed action (overlap) light baryon masses vs. $a^2$.}
\label{fig-baryons}
\end{center}
\end{figure}

In order to investigate the effects of the mixed action setup in other observables, we have computed the light baryon masses in the mixed action and the unitary setup. We have used smeared-smeared correlators with the same setup as in \cite{Alexandrou:2008tn}. The quark mass dependence of the nucleon and delta masses for one ensemble are depicted in Fig. \ref{fig-baryons} (left). At the matching mass (indicated by the vertical bar), the masses in both setups are compatible with each other, within statistical error. The situation is similar at a smaller lattice spacing, which is shown in the right panel of Fig. \ref{fig-baryons}. This indicates that the zero mode effects are much smaller in the baryon sector, at least in the case of light baryon masses. For more details about this analysis, we refer to an upcoming publication \cite{publ2}.

\section{Unitarity violations in the scalar correlator}
\begin{figure}[]
\begin{center}
\includegraphics[width=0.28\textwidth,angle=270]{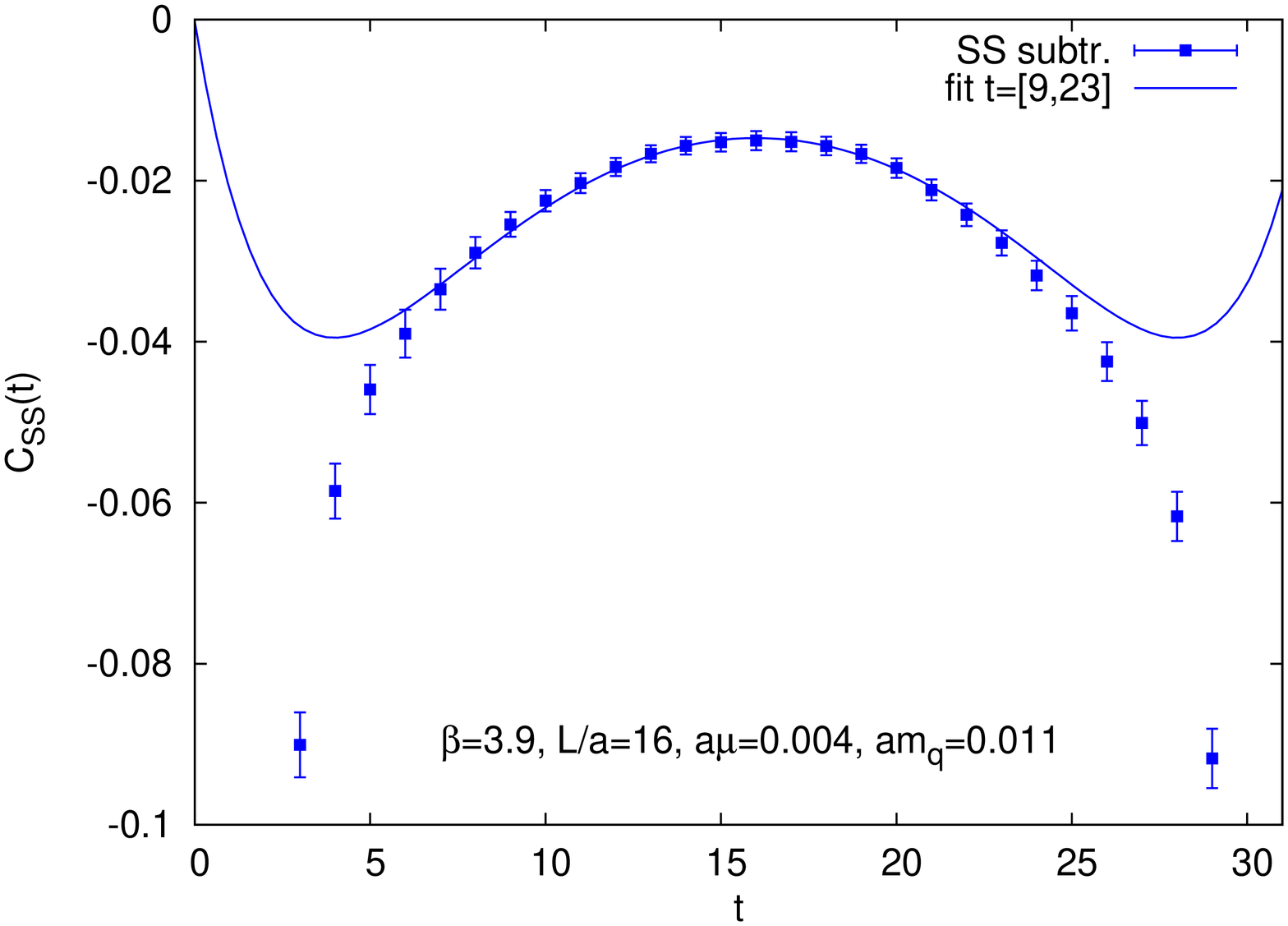}
\includegraphics[width=0.28\textwidth,angle=270]{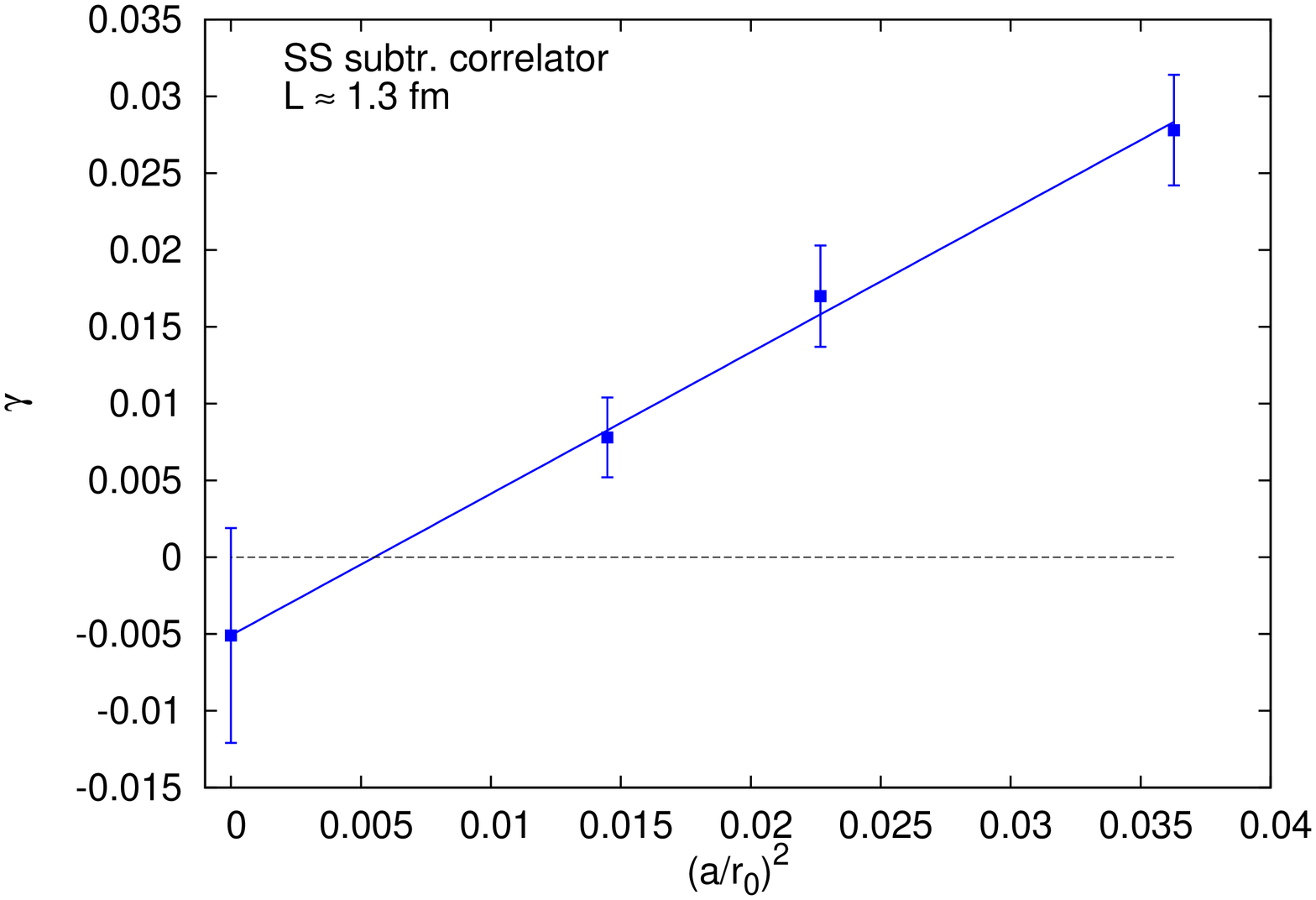}
\caption{(left) The SS subtr. correlation function at the matching mass . (right) Continuum limit scaling of the fitting parameter $\gamma$.}
\label{fig-scalar}
\end{center}
\end{figure}

The final aspect that we shortly discuss are the unitarity violations in the scalar correlator, present in the mixed action setup. As we have already stated, they can be minimized by matching the quark masses, but even at the matching mass they can not be entirely eliminated. One effect of this kind regards the scalar correlation function, which can obtain an unphysical negative contribution from one kind of diagrams. This effect in the mixed action setup has been analyzed within the framework of Partially Quenched Chiral Perturbation Theory in ref.~\cite{golterman05}. At large time $t$, the dominant contribution to the scalar correlation function is:
\begin{equation}
\label{ss-violation}
 C_{SS}(t)\stackrel{t\rightarrow\infty}{=}  -\frac{B_0^2}{2L^3}\frac{e^{-2M_{VV}t}}{M_{VV}^3}\gamma_{SS}a^2 t.
\end{equation} 
If we define $\gamma\equiv \frac{B_0^2\gamma_{SS}}{2(M_{VV}L)^3} a^2$ and take 
periodic boundary conditions in time, we obtain: $C_{SS}(t)\stackrel{t\rightarrow\infty}{=}
- \gamma \left(t\,e^{-2M_{VV}t}+(T-t)\,e^{-2M_{VV}(T-t)}\right)$.
In order to analyze this formula, we use the SS correlator with explicitly 
subtracted zero modes -- a field theoretically not clean procedure which we
use here only to have some estimate of the unitarity violation 
effect. We then extract $\gamma$ for each light-quark, small-volume ensemble, by 
fitting formula (\ref{ss-violation}) to the averaged SS correlator. An example of 
such fit is depicted in Fig. \ref{fig-scalar} (left). The right panel of this 
figure shows the extrapolation of $\gamma$ to the continuum limit. The linear 
dependence in $a^2$ and the value consistent with zero for $a=0$ is compatible 
with the hypothesis that the unitarity violations in the mixed setup are 
$\mathcal{O}(a^2)$ effects and can be described by formula (\ref{ss-violation}). 
For more details about this test, we again refer to \cite{publ2}.

\end{document}